\documentclass{jpsj-suppl}
\usepackage{txfonts} %Please comment out this line unless the txfonts package is availabe in your LaTeX system.
\usepackage{wrapfloat}
\usepackage{amssymb}

\def\Vec#1{\mbox{\boldmath $#1$}}
\def\cH{{\mathcal H}}
\def\infint{\int_{-\infty}^\infty}
\def\dstyle{\displaystyle}
\def\vS{{\Vec S}}  %(*** ƒxƒNƒgƒ‹ 'r ***)
\def\rB{{\mathrm B}}
\def\rd{{\mathrm d}}
\def\re{{\mathrm e}}

\title{Randomness Effect on the Temperature Dependence of the Finite Field Magnetization of a One-Dimensional Spin Gapped System}

\author{Kiyomi \textsc{Okamoto}}

\inst{Department of Physics, Tokyo Institute of Technology, 2-12-1 Ookayama, Meguro-ku, Tokyo 152-8551, Japan}

\email{kokamoto@phys.titech.ac.jp}

\recdate{Janurary 24, 2013}

\abst{We investigate the temperature dependence of the finite-field magnetization of the 
$S=1/2$ bond-alternating $XY$ model in
the Random magnetic field along the $z$-direction having the Lorentzian distribution.
The random-averaged free energy can be exactly calculated,
which enables us to obtain the thermodynamical quantities.
The temperature dependence of the finite-field magnetization shows various behaviors depending on the parameters $\delta$
and $\Gamma$, where $\delta$ and $\Gamma$ are the bond-alternation parameter and the Lorentzian distribution parameter, respectively.}

\kword{quantum spin chain, magnetization, random magnetic field}

\begin{document}
\maketitle

\section{Introduction}
%The spin gapped systems
Many one-dimensional antiferromagnets have singlet ground states
with the energy gaps in the excitation spectra.
A typical example is the simple $S=1$ chain having the Haldane ground state \cite{haldane,affleck}.
Another example is the bond-alternating $S=1/2$ chain having the dimer ground state \cite{bray}.
When the magnetic field applied to the spin-gapped system is so strong that
the Zeeman energy exceeds the energy gap between the singlet ground state and the lowest state with finite magnetization,
the system becomes gapless and can be described by the Tomonaga-Luttinger liquid theory.
In this case,
the low energy properties of the system are thought to be very similar to those of
simple antiferromagnetic $S=1/2$ chain in zero magnetic field which is gapless.
Thus, one may think that
the temperature dependence of the magnetization may be
very similar to that of the Bonner-Fisher's magnetic susceptibility \cite{bonner-fisher}.
However, as far as the temperature dependence of magnetization is concerned,
the above naive expectation does not hold,
as explained in the following.

Hida, Imada and Ishikawa \cite{hida} studied the quantum sine-Gordon model with finite winding number
which can be related to the magnetization of spin chains.
The $S=1/2$ two-leg ladder was numerically studied
by Wang and Yu \cite{wang-yu} by use of the transfer-matrix renormalization
group method,
and also by Wessel, Olshanii and Haas \cite{wessel} by use of the quantum Monte Carlo method.
Both groups found a minimum-maximum (Min-Max) behavior (similar to that of Fig.\ref{fig:mag}(a))
of the magnetization as a function of the temperature,
which is different from the Bonner-Fisher's behavior.
Honda {\it et al.} \cite{honda} measured the magnetization of ${\rm Ni(C_5 H_{14} N_2)_2 N_3(PF_6)}$ 
(often abbreviated as NDMAP) and observed the Min-Max behavior.
Maeda, Hotta and
\hyphenation{Oshikawa}
Oshikawa (MHO) \cite{maeda}
performed the quantum Monte Carlo calculation for the $S=1$ spin chain
and found the Min-Max behavior.
MHO also clarified the mechanism and the universal nature of the Min-Max behavior.
After MHO,
the Min-Max behavior was found in several cases \cite{gong,ding,bouillot,ninios}.
The Min-Max behavior was also found
in the case that the origin of the spin gap is the spontaneous symmetry breaking \cite{suga}.

Thus, the Min-Max behavior of the one-dimensional spin-gapped systems
seems to be established.
However, the effect of the randomness on the Min-Max behavior
has not been studied so far.
Since the Min-Max behavior is based on the subtle balance
between the distribution function and the density of states (DOS), 
as was explained by MHO,
we should treat the randomness effects very carefully.
In this paper,
we study the effect of randomness on the Min-Max behavior
of the bond-alternating $S=1/2$ $XY$ model in the random
magnetic field with the Lorentzian distribution
which is exactly solvable.

%********************************************************************
\section{Model}
%********************************************************************
%\subsection{Subsection}
%\subsubsection{Subsubsection}
We investigate the model
\begin{equation}
    \cH
    = J\sum_{j=1}^{2N} [1+(-1)^j \delta](S_j^x S_{j+1}^x + S_j^y S_{j+1}^y)
      - g\mu_\rB\sum_{j=1}^{2N} (H + H_j) S_j^z, 
\end{equation}
where $\vS$ is the spin-1/2 operator,
$J>0$ is the coupling constant between neighboring spins,
$2N$ is the number of spins,
$\delta$ is the bond-alternation parameter ($0 \le \delta \le 1$)
and $H$ is the uniform magnetic field along the $z$ direction.
The quantity $H_j$ is the random magnetic field along the $z$ direction
with the Lorentzian distribution
\begin{equation}
    P(H_j)
    = {1 \over \pi} {\Gamma \over H_j^2 + \Gamma^2}.
\end{equation}
We suppose that there is no correlation between $H_i$ and $H_j$ when $i \ne j$.
Hereafter we set $J=1$ (unit energy) and also $k_\rB = 1$, $\hbar = 1$ and $g\mu_\rB = 1$.

The exact solvability of this model in the $\delta=0,~\Gamma = 0$ case was shown by  
Lieb, Schultz and Mattis \cite{lieb},
and that in the $\delta>0,~\Gamma = 0$ case by Pincus \cite{pincus}.
Nishimori \cite{nishimori} exactly solved the $\delta=0,~\Gamma > 0$ case by use of
the Lloyd method \cite{lloyd}.
The full case $\delta>0,~\Gamma > 0$ was exactly solved by the present author \cite{okamoto1990}.

%********************************************************************
\section{Behavior of the Magnetization as a Function of Temperature}
%********************************************************************

Here we summarize the results of my previous paper\cite{okamoto1990}.
The random-averaged free energy per one spin can be exactly calculated as
\begin{equation}
    \tilde F
    \equiv {F \over 2N}
    = - T \infint \rd \omega\,\bar\rho_H(\omega){1 \over 1 + \re^{\omega/T}} + {H \over 2},
\end{equation}
where $T$ is the temperature.
The quantity $\bar\rho_H(\omega)$ is the DOS of Jordan-Wigner fermions given by
\begin{equation}
    \bar\rho_H(\omega) = \bar\rho_0(\omega_H),~~~~~
    \omega_H \equiv \omega + H,
\end{equation}
with
\begin{equation}
    \bar\rho_0(\omega)
    = {1 \over \pi\sqrt{AB}}\sin{\theta+\varphi \over 2},
\end{equation}
where $A$, $B$, $\theta$ and $\varphi$ are defined by
\begin{eqnarray}
   &&A \equiv \sqrt{[(\omega + \delta)^2 + \Gamma^2][(\omega - \delta )^2 + \Gamma^2]
                      \over \omega^2 + \Gamma^2},~~~~~ 
   B \equiv \sqrt{[(\omega + 1)^2 + \Gamma^2][(\omega - 1)^2 + \Gamma^2]
                      \over \omega^2 + \Gamma^2}, \\
   &&\tan\theta \equiv 
                {\Gamma(\omega^2 + \Gamma^2 + \delta^2 ) \over \omega(\omega^2 + \Gamma^2 - \delta^2 )},~~~~~
   \tan\varphi \equiv 
                {\Gamma(\omega^2 + \Gamma^2 + 1) \over \omega(\omega^2 + \Gamma^2 - 1)},
                ~~~~~~~~(0 \le \theta,\,\varphi \le \pi),  
\end{eqnarray}
\begin{wrapfigure}[14]{r}{7cm}
   \scalebox{0.3}{\includegraphics{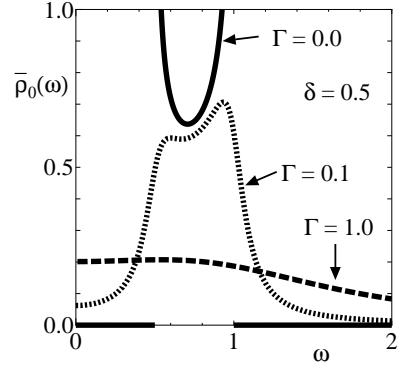}}
   \caption{DOS of the Jordan-Wigner fermions as a function of $\omega$
            for the $\Gamma = 0,0,\,0.5,\,1.0$ cases when $\delta = 0.5$ and $H=0$.
            }
   \label{fig:dos}
\end{wrapfigure}
respectively.
As is shown in Fig.\ref{fig:dos},
the DOS $\bar\rho_0(\omega)$ is everywhere non-zero when $\Gamma>0$.
In the absence of the random field ($\Gamma=0$),
we obtain
\begin{equation}
    \rho_0^{(0)}(\omega)
    = \begin{cases}
      \dstyle{{1 \over \pi}{\omega \over \sqrt{(\omega^2-\delta^2)(1-\omega^2)}}} &(\delta < |\omega| <1) \\
       0 &\text{(otherwise)}.
      \end{cases}
\end{equation}
On the other hand,
in the absence of the bond alternation ($\delta = 0$),
the DOS $\bar\rho_0(\omega)$ 
is reduced to Nishimori's result \cite{nishimori}.

\begin{figure}[h]
   \begin{center}
%   \scalebox{0.25}{\includegraphics{d05g00h06-paper.eps}}~~~
   \scalebox{0.25}{\includegraphics{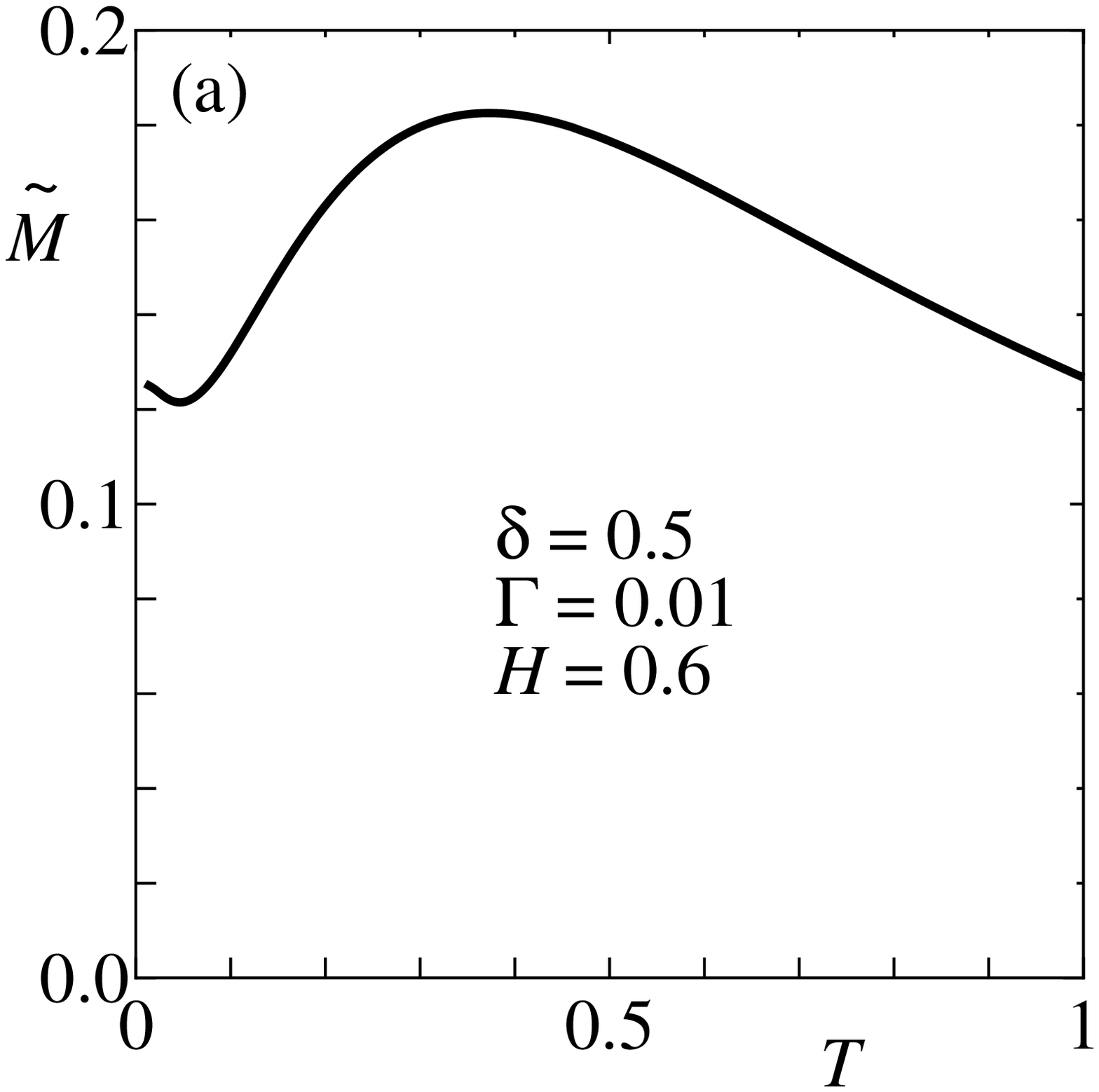}}~~~
   \scalebox{0.25}{\includegraphics{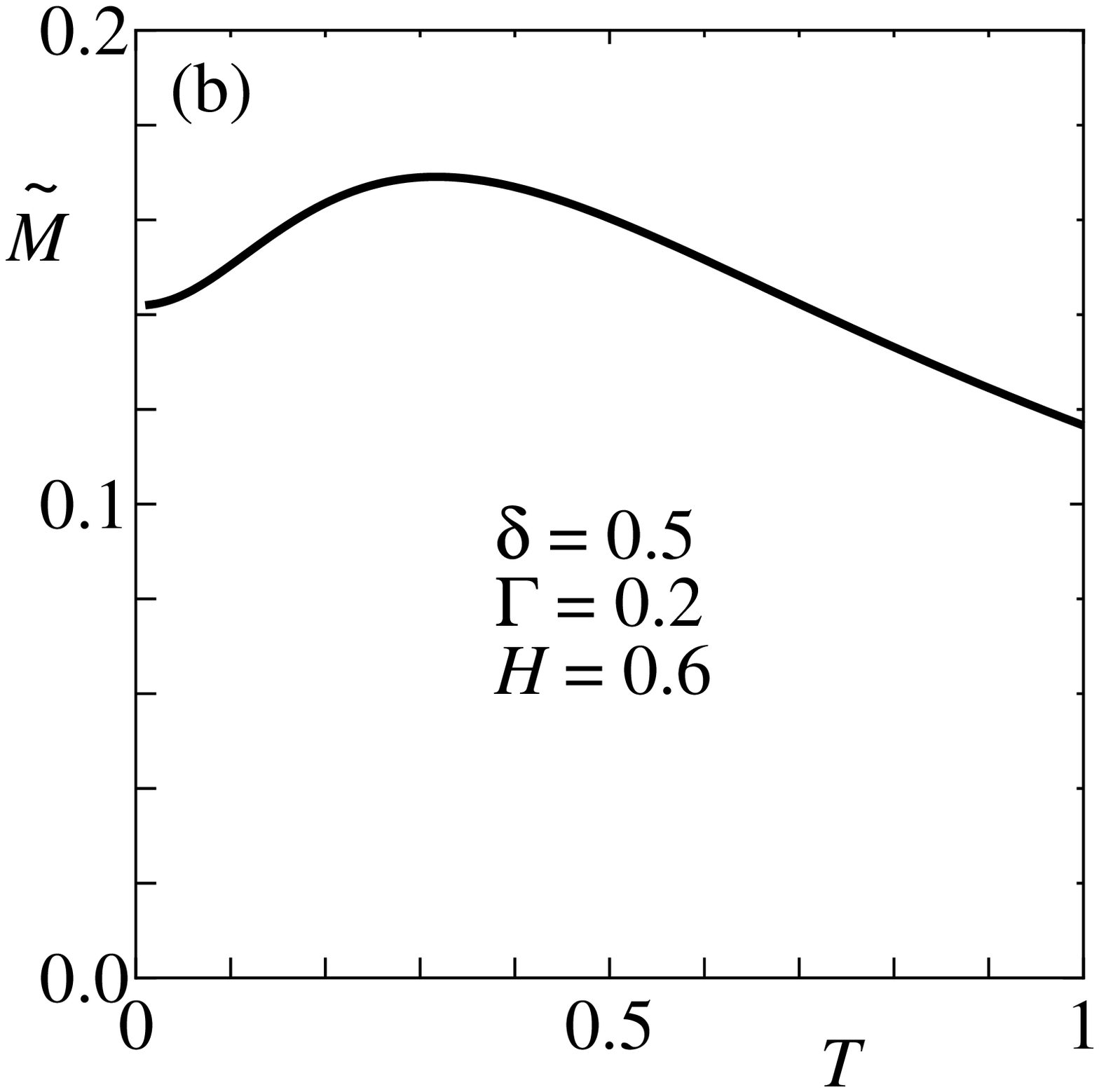}}~~~
   \scalebox{0.25}{\includegraphics{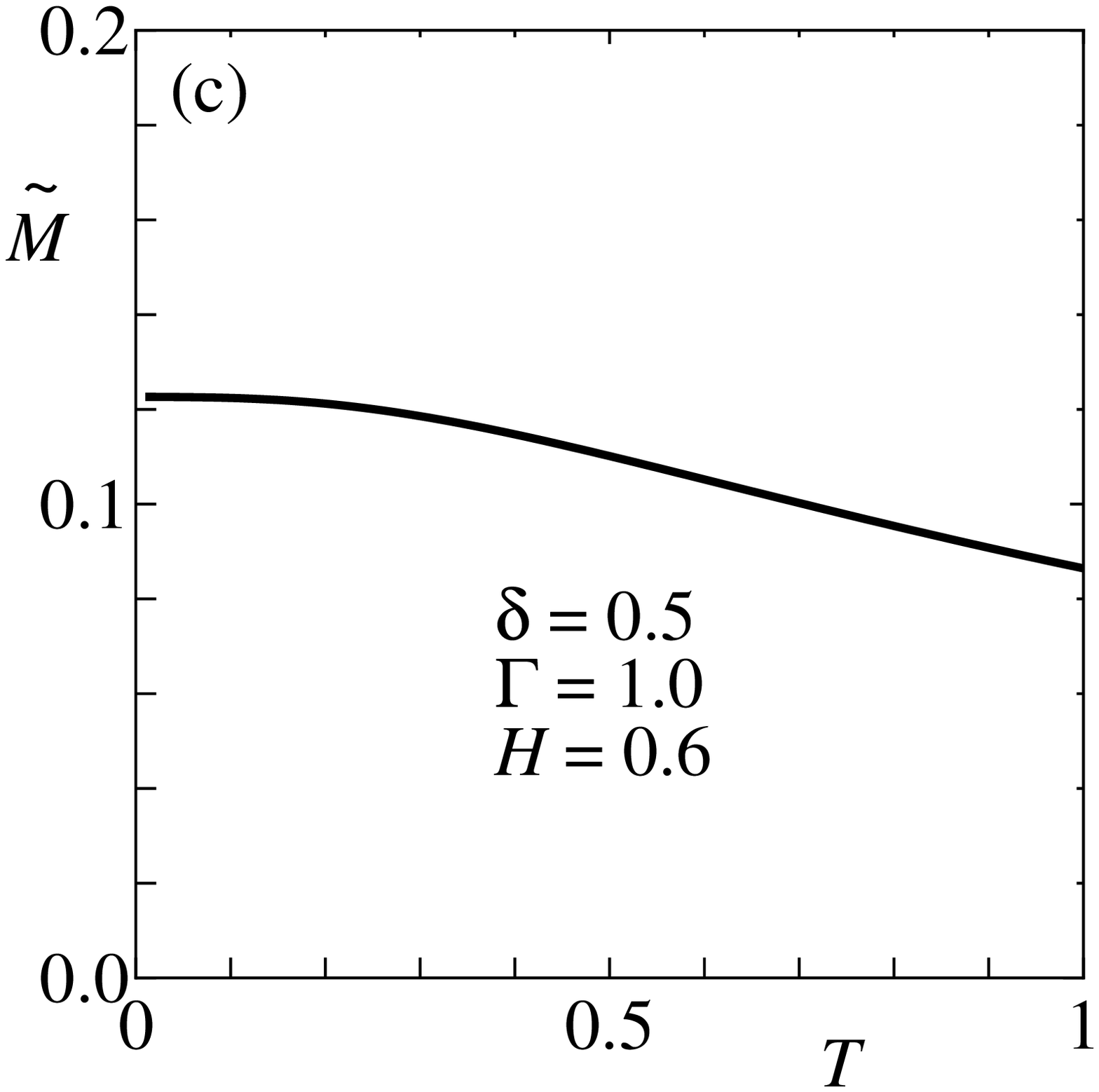}}
   \end{center}
   \caption{Typical behaviors of the magnetization per one spin $\tilde M \equiv M/2N$ as functions of $T$
            for the (a) $\Gamma = 0.01$, (b) $\Gamma = 0.2$ and (c) $\Gamma = 1.0$ cases when $\delta = 0.5$ and $H=0.6$.
            The behavior of $\tilde M$ in the $\Gamma = 0$ case is very similar to that in the $\Gamma = 0.01$ case.}
   \label{fig:mag}
\end{figure}

Once $\tilde F$ is known,
it is easy to calculate the magnetization in the $z$-direction per one spin as
\begin{equation}
    \tilde M
    \equiv {M \over 2N}
    = \infint \rd \omega \,\bar\rho_H(\omega) {1 \over 1 + \re^{\omega/T}} - {1 \over 2}.
    \label{eq:M}
\end{equation}
Typical behaviors of $\tilde M$ as functions of $T$ are shown in Fig.\ref{fig:mag}.
Figure 3 shows the behavior patterns of $\tilde M$ on the $\Gamma-H$ plane
which was obtained by the numerical analyses of eq.(\ref{eq:M}).
\begin{figure}[h]
   \begin{center}
   \scalebox{0.3}{\includegraphics{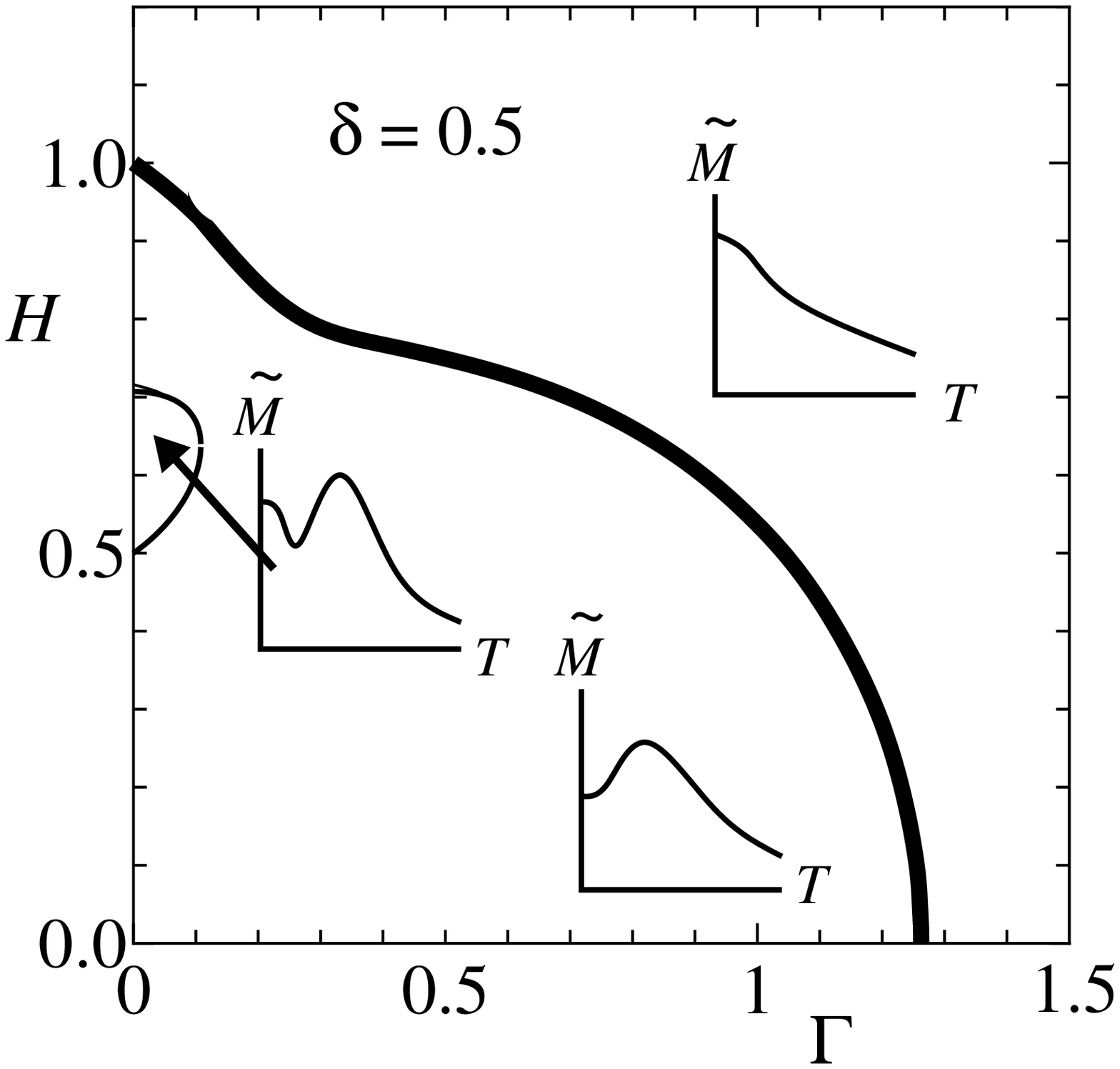}}~~~
   \scalebox{0.3}{\includegraphics{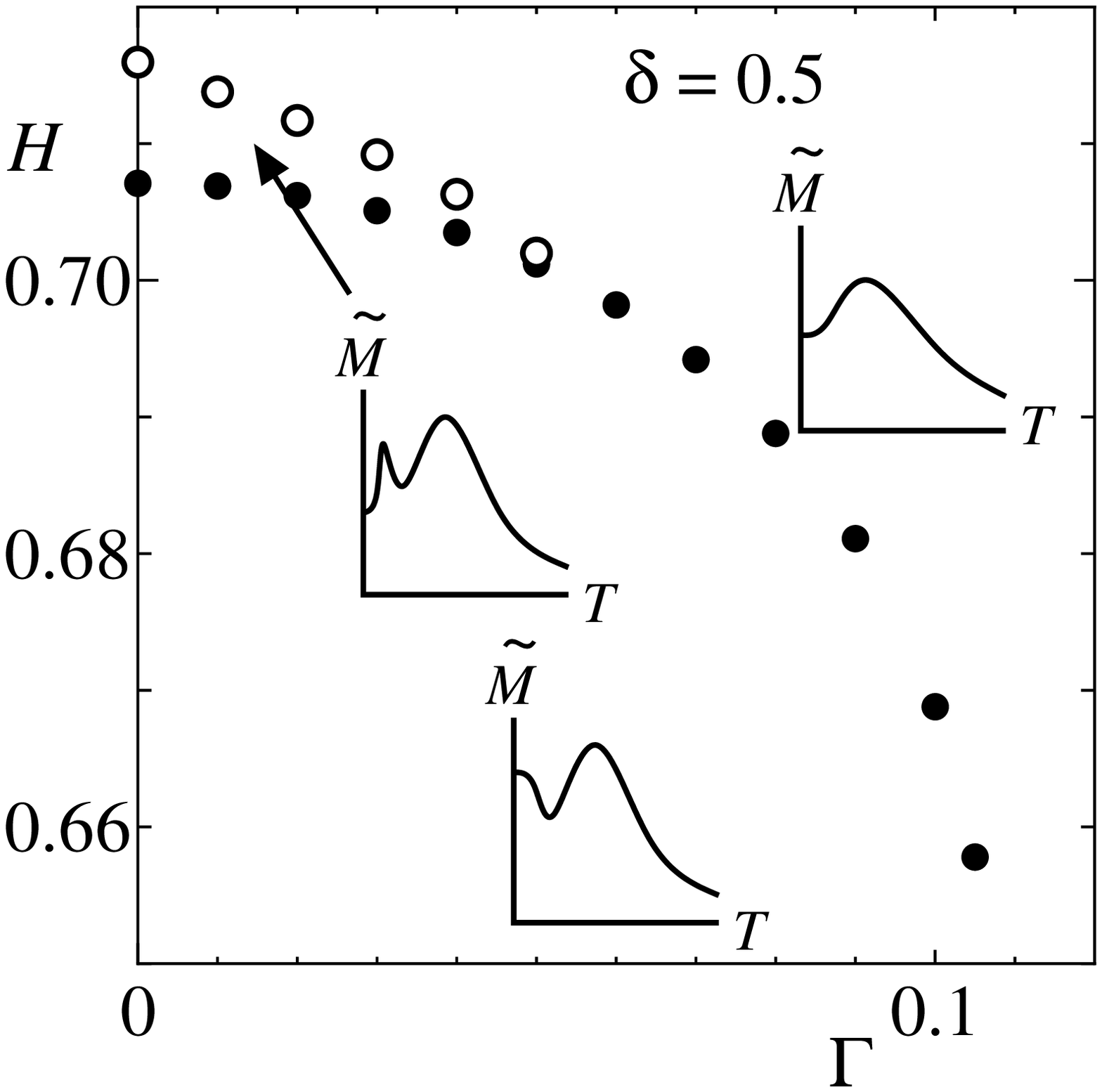}}
   \end{center}
   \caption{Patterns of behavior of $\tilde M$ on the $\Gamma-H$ plane when $\delta = 0.5$.
            The details near $H = 0.7$ are shown in the right panel.}
   \label{fig:phase}
\end{figure}

%********************************************************************
\section{Discussion}
%********************************************************************

At low temperatures,
by use of the Sommerfeld expansion \cite{sommerfeld},
the magnetization $\tilde M$ can be expanded as
\begin{equation}
    \tilde M
    = -{1 \over 2}
      + \int_{-\infty}^H \rd \omega\,\bar\rho_0(\omega) 
      + {\pi^2 {{\bar\rho}_0}'(H) \over 6}T^2
      + {7\pi^4 {{\bar\rho}_0}'''(H) \over 360}T^4
      + O(T^6),
    \label{eq:sommerfeld}
\end{equation}
where ${{\bar\rho}_0}'(H) = d{\bar\rho}_0(\omega)/d\omega|_{\omega = H}$
and  ${{\bar\rho}_0}'''(H) = d^3{\bar\rho}_0(\omega)/d\omega^3|_{\omega = H}$.
Therefore the initial change of $\tilde M$
is positive or negative according as
$\bar{\rho_0}'(H) > $ or $\bar{\rho_0}'(H) < 0$.
As far as we take the terms up to $T^4$ in the rhs of eq.(\ref{eq:sommerfeld}),
the condition for the Min-Max behavior like Fig.\ref{fig:mag}(a)
is ${{\bar\rho}_0}'(H) < 0$ and ${{\bar\rho}_0}'''(H) > 0$.
When $\delta = 0.5$ and $\Gamma = 0$,
the Min-Max behavior of $\tilde M$ is realized for $0.5 < H < 1/\sqrt{2}$.
This behavior still holds even in the presence of small $\Gamma$ as shown in Fig.\ref{fig:phase}.
Thus the shape of $\bar\rho_0(\omega)$ in Fig.\ref{fig:dos} and
the patterns of the behavior of $\tilde M$ in Fig.\ref{fig:phase}
are consistent with each other.
When $\Gamma>0$, 
the excitation gap vanishes and, as a result,
the divergence of the DOS also vanishes.
Thus the existence of excitation gap and divergence of the DOS
are not the necessary condition for
the Min-Max behavior of $\tilde M$ like Fig.\ref{fig:mag}(a).
Roughly speaking, the sharp double peaks of the DOS would rather be important.

In conclusion, 
we have investigated the effect of the randomness on the Min-Max behavior
for the first time.
For the specific heat $C$
we also observed interesting behaviors,
for instance,
the double peak behavior with a shoulder.
Their details will be discussed elsewhere.

\section*{Acknowledgment}
This work was
partly supported by grants-in-aid for Scientific Research (B)
(No.23340109) and Scientific Research (C) (No. 23540388), 
from the Ministry of Education, Culture, Sports, Science,
and Technology of Japan.


\begin{thebibliography}{99}

\bibitem{haldane}
F. D. M. Haldane: Phys. Lett. \textbf{93A} (1983) 464.

\bibitem{affleck}
for a review, I. Affleck: J. Phys: Cond. Matt. \textbf{1} (1989) 3047.

\bibitem{bray}
for a review, J. W. Bray, L. V. Interrante, I. S. Jaobs, and J. C. Bonner:
\textit{Extended Linear Chain Coumpounds}, ed. J. S. Miller
(Prenum Press, New Yoirk, 1983) Vol.3.

\bibitem{bonner-fisher}
J. Bonner and M. E. Fisher: Phys. Rev. \textbf{135} (1964) A640.

\bibitem{hida}
K. Hida, M. Imada, and M. Ishikawa: J. Phys. C \textbf{16} (1983) 4945.

\bibitem{wang-yu} 
X. Wang and Lu Yu:
Phys. Rev. Lett. \textbf{84} (2000) 5399.

\bibitem{wessel} 
S. Wessel, M Olshanii, and S. Haas:
Phys. Rev. Lett. \textbf{87} (2001) 206407.

\bibitem{honda} 
Z. Honda, K. Katsumata, Y. Nishimayma, and I. Harada:
Phys. Rev. B \textbf{69} (2001) 064420.

\bibitem{maeda} 
Y. Maeda, C. Hotta and M. Oshikawa:
Phys. Rev. Lett. \textbf{99} (2007) 057205.

\bibitem{gong} 
S.-S. Gong, S. Gao, and G. Su:
Phys. Rev. B \textbf{80} (2009) 014413.

\bibitem{ding}
L. J. Ding, K. L. Yao, and H. H. Fua:
Phys. Chem. Chem. Phys. \textbf{13} (2011) 328.

\bibitem{bouillot}
P. Bouillot, C. Kollath, A. M. L\"auchli, M. Zvonarev, B. Thielemann, C. R\"uegg,
E. Orignac, R. Citro, M.~Klanj\v{s}ek, C. Berthier, M. Horvati\'c, and T. Giamarchi:
%P. Bouillot \textit{et. al}:
Phys. Rev. B \textbf{83} (2011) 054407.

\bibitem{ninios}
K. Ninios, T. Hong, T. Manabe, C. Hotta, S. N. Herringer, M. M. Turnbull, C. P. Landee,
Y. Takano, and H. B. Chan
%K. Ninios \textit{et. al}:
Phys. Rev. Lett. \textbf{108} (2012) 097201.

\bibitem{suga} 
S. Suga:
J. Phys. Soc. Jpn. \textbf{77} (2008) 074717.

\bibitem{lieb} 
E. Lieb, T. Schultz, and D. Mattis:
Ann. Phys. (NY) \textbf{16} (1961) 407.

\bibitem{pincus}
P. Pincus:
Solid State Commun. \textbf{9} (1971) 1971.

\bibitem{nishimori} 
H. Nishimori: Phys. Lett. \textbf{100A} (1984) 239.

\bibitem{lloyd} 
P. Lloyd:
J. Phys. C \textbf{2} (1969) 1717.

\bibitem{okamoto1990}
K. Okamoto: J. Phys. Soc. Jpn. \textbf{59} (1990) 4286.


\bibitem{sommerfeld}
for instance, N. W. Ashcroft and N. D. Mermin:
\textit{Solid State Physics} 
(Thomson Learning, London, 1976) p. 760.

\end{thebibliography}
\end{document}